\documentclass{ws-procs9x6}

\begin{document}

\title{Relation between the chiral and deconfinement phase transitions}

\author{Y. Hatta}

\address{RIKEN BNL Research Center,\\
Brookhaven National Laboratory, \\
Upton, NY 11973, USA\\
 E-mail: hatta@quark.phy.bnl.gov}

\maketitle

\abstracts{ Lattice QCD simulations at finite temperature have
shown that the chiral phase transition in the chiral limit and the
deconfinement phase transition in the quenched limit are
continuously connected. I emphasize the nontriviality of this
result and propose an unconventional  scenario which naturally
explains the existing lattice data. The continuity of the two
phase transitions is a manifestation of the familiar
glueball-meson mixing, which can be traced back to the properties
of QCD at zero temperature.}

\section{Introduction and motivation}
In QCD at finite temperature, there are two kinds of phase
transitions in two different limits of the quark mass parameter.
The chiral phase transition in the chiral limit and the
deconfinement phase transition in the quenched limit. Away from
these limits, the meanings of these phase transitions become less
transparent. Nevertheless, the notions of deconfinement and chiral
symmetry restoration are indispensable for our understanding of
 the quark-gluon plasma phase. In this talk we discuss the
 possible deep relation between the two phase transitions in the intermediate
 quark mass region.

In Table 1, some key properties of the two phase transitions are
listed.

\begin{table}[ph]
\tbl{QCD phase transitions: defining properties.} {\footnotesize
\begin{tabular}{@{}crrrr@{}}
\hline
{} &{} &{} &{} &{}\\[-1.5ex]
{} & chiral phase transition  & deconfinement phase transition\\[1ex]
\hline
{} &{} &{} &{} &{}\\[-1.5ex]
quark mass & 0 & $\infty$ \\[1ex]
symmetry & chiral symmetry & center symmetry \\[1ex]
order parameter & quark condensate & Polyakov loop \\[1ex]
\hline
\end{tabular}\label{table2} }
\vspace*{-13pt}
\end{table}

At first sight, it does not make sense to talk about the
relationship between the two phase transitions. They are defined
in completely different theories in the first place. Symmetries
are different, and the order parameters are different. Just by
looking at the table, little do we suspect that the two phase
transitions have anything to do with each other. Therefore, it is
quite natural that they have long been considered as distinct
phase transitions and questions like ``Which phase transition
occurs first?" have been addressed many times in the literature.

However, putting these theoretical speculations aside, finite
temperature lattice simulations have repeatedly shown that the two
phase transitions occur at the {\it same} critical temperature.
Fig 1. is the most commonly accepted phase diagram of QCD in the
temperature-quark mass plane.\cite{karsch} This figure shows that,
for all values of the quark mass, there is a single (crossover)
phase transition which smoothly connects the chiral phase
transition in the chiral limit and the deconfinement phase
transition in the quenched theory.

\begin{figure}[ht]
\centerline{\epsfxsize=2.7in\epsfbox{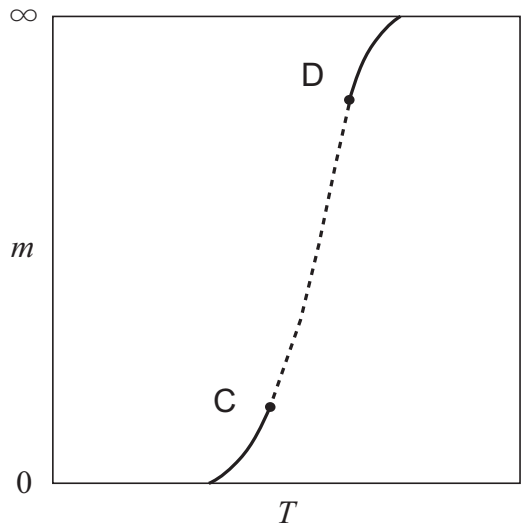}} \caption{The
QCD phase diagram in the temperature-quark mass plane. $T$ is the
temperature and $m$ is the quark mass.\label{inter} Solid (dotted)
lines represent first order (crossover) phase transition.
$\textsf{C}$ and $\textsf{D}$ are second order phase transition
points.}
\end{figure}

We emphasize that this is a very nontrivial result. Theorists were
able to predict the order of phase transitions at $m=0$ and
$m=\infty$.\cite{yaffe,rob} Theorists could also predict that the
first order chiral (deconfinement) phase transition would turn
into a crossover if the quark mass was increased from 0 (decreased
from $\infty$). However, no one could predict the {\it global}
structure of the phase diagram shown in Fig. 1 because {\it
physics at the intermediate quark mass region is non-universal}.
One cannot invoke the usual universality argument of phase
transitions to predict anything in this region. The continuity of
the two phase transitions is a consequence of the non-perturbative
dynamical effect of QCD, which is far from obvious. And, of
course, such a {\it non-}universal phenomenon is one of  the most
interesting aspects of a given theory. In this report, we will
propose a novel scenario\cite{link} which naturally explains the
puzzling lattice data, Fig. 1.

\section{The level repulsion scenario}
The interplay between the two transitions is an old but
interesting problem.  First, Gocksch and Ogilvie
observed\cite{gock} that what is responsible for the breaking of
center symmetry is (the inverse of) the {\it constituent} quark
mass rather than the current quark mass. This suggests that the
Polyakov loop and the chiral dynamics are closely coupled. See,
also, Ref.\cite{satz} for more recent works. So far, most of the
works on this subject focused on the coupling of the Polyakov loop
and the sigma field. (See, however, Ref.\cite{red}.) Here we point
out that, in the presence of dynamical quarks, the deconfinement
phase transition can equivalently be characterized in terms of the
{\it glueballs}. Specifically, we have predicted\cite{link}  that
the screening mass of the $0^+$ electric\footnote{See, for
example, Ref.\cite{datta} for the meaning of these quantum
numbers. 'Electric' means that the glueball interpolating operator
contains $A_0$'s or timelike links. One can also consider $0^+$
'magnetic' glueballs (composed only from spacelike links) which,
in principle, mix with the electric ones. However, near the
critical temperature and above, it has been observed\cite{hart}
that the mixing is very weak, could be absent.} glueball
 goes to zero with the specified critical exponent at the second order deconfinement
 phase transition.\cite{sa}
In the case of color SU(2), this screening state is responsible
for the weak divergence of the specific heat at the Z(2) phase
transition but is distinct from the true order parameter field,
namely, the Polyakov loop. However, in the case of color SU(3), at
the point \textsf{D}, nonzero expectation value of the Polyakov
loop induces mixing between the glueball and the Polaykov loop.
Therefore, the glueball is an equivalent critical field at
\textsf{D}.\footnote{Such a mixing always takes place at generic
end-points. See, Ref.\cite{hatta} for an explicit example in the
case of the end-point at finite density.} Datta and Gupta observed
a significant decrease of the $0^+$ glueball screening mass near
the SU(2) phase transition in the quenched simulation.\cite{datta}
We believe that the mass will  go to zero in the infinite lattice
volume limit.

Next we observe that the glueball field $G$ must mix with the
sigma field $\sigma$
 so that the correct massless field at \textsf{D} is a linear combination of the two;
\begin{eqnarray}
\phi=G\cos \theta  + \sigma\sin \theta,\ \ \ \ \  \sin \theta
\approx 0.\label{g}
\end{eqnarray}
 The orthogonal linear
combination with large sigma field content,
\begin{eqnarray}
\phi'=-G\sin \theta + \sigma \cos \theta,
\end{eqnarray}
is massive. Now the key question is the behavior of the mixing
angle $\theta$ as the quark mass varies. If the mixing angle
remains small at small values of the quark mass, the $\phi'$
field, which is massive at \textsf{D}, would become massless at
\textsf{C} because the critical field at \textsf{C} is dominantly
sigma-like.\cite{gavin} However, if this were the case, the
coincidence of the two critical temperatures for all values of the
quark mass would be a pure accident. Fig. 1 is most naturally
explained by postulating that the critical field at \textsf{C} is
again the $\phi$ field. Namely, the two second order
 phase transitions at \textsf{C} and \textsf{D} are driven by the {\it same} field.
 This is possible only if the mixing angle changes from $\theta \approx 0$ to $ \approx
\pi/2$.\footnote{The mixing angle will go to exactly $\pi/2$
 at the chiral symmetry restoration point in the chiral limit.}
 Such a continuous variation of the mixing angle is typical of a {\it level repulsion} in quantum
 mechanics. Thus we have arrived at a novel scenario of the finite temperature QCD phase transition:
   Due to a level repulsion between the $\phi$ and the $\phi'$ fields, the $\phi$ field continues
  to be the lightest screening state for all values of the quark mass. Simultaneous divergences and peaks
  in various susceptibilities are simply caused by the dropping of the $\phi$ field screening mass.

Moreover, we conjecture that this scenario is realized at all
temperatures, not only near the critical temperature. The level
repulsion between the scalar glueball and the sigma meson takes
place already at zero temperature as one can easily convince
oneself by considering the scalar meson and glueball mass spectrum
at zero temperature.\footnote{We assume that, in accordance with
popular belief, the lightest glueball mass in the real world does
not change appreciably from its value in the quenched theory.}
Therefore, our scenario can be naturally embedded in the entire
phase diagram, Fig. 1. Note that this argument is made possible
only when one characterizes the deconfinement phase transition in
terms of the glueball. (The Polyakov loop cannot be defined at
zero temperature.)

\section{Conclusion}
We have pointed out the importance of the glueball screening
states for the understanding of the QCD phase diagram  for all
values of the quark mass and the temperature.
 Compared to Polyakov loops, glueballs have been much less studied
  on a lattice at finite temperature with or without dynamical
  quarks.\cite{datta,hart,gavai,ishii} We expect that further glueball
measurements will
 provide rich information
  on the nature of the thermal QCD phase transition. On the other hand, it turned out to be
  very difficult to
reproduce the lattice result, Fig. 1, in model calculations. The
main difficulty is to keep track of the important coupling between
the two fields in the intermediate quark mass region where there
is no guiding principle (no symmetry) to construct the effective
potential. (For details and discussions, see Ref.\cite{fuku2}.)
This remains to be a theoretical challenge.

\section*{Acknowledgments}
This work has been done in collaboration with Kenji Fukushima. I
thank the organizers of Strong and Electroweak Matter 2004 for the
stimulating meeting.
\appendix

\end{document}